\documentclass[twocolumn,prl,showpacs]{revtex4}
\usepackage{graphicx}
\usepackage{dcolumn}
\usepackage{bm}
\begin{document}
\title{Ring Exchange Mechanism for Triplet Superconductivity \\
in a Two-Chain Hubbard Model: \\
Possible Relevance to Bechgaard Salts}
\author{Y. Ohta,$^{1,2}$ S. Nishimoto,$^{3}$ 
T. Shirakawa,$^{2}$ and Y. Yamaguchi$^{2}$}
\affiliation{$^{1}$Department of Physics, Chiba University, 
Chiba 263-8522, Japan} 
\affiliation{$^{2}$Graduate School of Science and Technology, 
Chiba University, Chiba 263-8522, Japan}
\affiliation{$^{3}$Institut f\"ur Theoretische Physik, Universit\"at 
G\"ottingen, D-37077 G\"ottingen, Germany} 
\date{23 December 2004}
\begin{abstract}
The density-matrix renormalization group method is used 
to study the ground state of the two-chain zigzag-bond 
Hubbard model at quarter filling.  
We show that, with a proper choice of the signs of 
hopping integrals, the ring exchange mechanism yields 
ferromagnetic spin correlations between interchain 
neighboring sites, and produces the attractive 
interaction between electrons as well as the long-range 
pair correlations in the spin-triplet channel, thereby 
leading the system to triplet superconductivity.  
We argue that this novel mechanism may have possible 
relevance to observed superconductivity in Bechgaard 
salts.  
\end{abstract}
\maketitle

More than two decades have passed since the discovery of 
superconductivity in so-called Bechgaard salts such as 
(TMTSF)$_2$X with X=PF$_6$, ClO$_4$, etc. 
\cite{jerome,bechgaard}.  
The mechanism of superconductivity of this strongly 
correlated quasi-one-dimensional (1D) electron system 
\cite{review} is, however, still an open issue.  
Recently, experimental evidences have been accumulating 
that the Cooper pairs of this system are in the 
spin-triplet state: they are from the measurements 
of the temperature dependence of Knight shift and 
spin-lattice relaxation ratio \cite{takigawa,lee1,lee2} 
as well as that of the upper critical field 
\cite{lee3,lee4,oh}.  

A variety of theoretical approaches have so far 
been made on this unconventional superconductivity 
\cite{lebed,tanaka,kuroki1,fuseya,shimahara,kuroki2}, 
but none of them is based upon real-space pairing 
mechanism possibly relevant due to their strong 
electron correlations.  
In this paper, we want to propose a novel mechanism 
that may have possible relevance to triplet 
superconductivity of this system.  We adopt a 
numerical approach in order to take fully into 
account the strong electron correlations.  

First, let us point out that the hopping integrals 
of (TMTSF)$_2$X have the unique structure as shown 
in Fig.~1(a) \cite{ducasse}; they show an alternating 
sign change along the zigzag bonds connecting two 
chains, while the sign along the 1D chain is always 
positive in the electron notation (i.e., for the bands 
of 3/4 filling of electrons).  We hereafter use the 
hole notation for convenience; i.e., the system is 
in the quarter filling of holes.  The signs of 
hopping integrals are then always negative along 
the 1D chain and are changing signs along the zigzag 
bonds.  

Let us then notice that, in Hubbard models defined 
on such triangular-lattice related structures with 
proper signs of hopping integrals, the ring exchange 
mechanism of two spins on a triangle makes the 
system a ferromagnet with full spin polarization 
if sufficiently large Hubbard interaction $U$ 
acts \cite{tasaki,penc,daul,noack}.  
The structure of hopping integrals of 
(TMTSF)$_2$X \cite{ducasse}, in fact, satisfies 
the ferromagnetic sign rule $t_1t_2t_3>0$ 
(in the hole notation) for three hopping integrals 
of all the triangles.  

Our mechanism for triplet superconductivity relies 
on this ferromagnetic coupling: spin-triplet coupling 
for ferromagnetism is `relaxed' to short-range 
spin-triplet correlations in realistic strength of 
$U$, thereby leading the system to a metallic state 
with ferromagnetic spin correlations.  
We then ask what this state actually is.  
\begin{figure}[t]
\begin{center}
\includegraphics[width=6.0cm,clip]{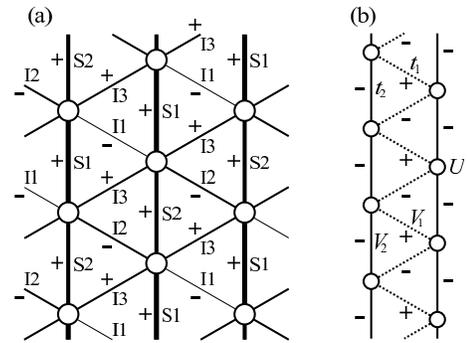}
\caption{Schematic representation of (a) the structure 
of hopping integrals for (TMTSF)$_2$X \cite{ducasse} 
and (b) the two-chain Hubbard model we use.  The signs 
of hopping integrals are shown in the electron notation 
in (a) and hole notation in (b).  
The parameter values for (TMTSF)$_2$ClO$_4$ 
at low temperatures, e.g., are: 
$t_{\rm S1}=287$, 
$t_{\rm S2}=266$, 
$t_{\rm I1}=-34.0$, 
$t_{\rm I2}=-64.1$, and 
$t_{\rm I3}=46.2$ meV \cite{ducasse}.}
\end{center}
\label{fig1}
\end{figure}

We calculate the ground state of the relevant 
two-chain Hubbard model by the density-matrix 
renormalization group (DMRG) method \cite{white} and 
show that indeed the system is metallic with 
ferromagnetic spin correlations.  Furthermore, we show 
that the attractive interaction acts between holes, 
being caused by the gain in kinetic energy due to 
ring exchange of holes.  
We also show that the superconducting pair correlations 
in the spin-triplet channel extend long-ranged in 
power-law length dependence, while the pair correlations 
in the singlet channel as well as the spin-density-wave 
(SDW) correlations decay exponentially, indicating that 
the system is in the state of spin-triplet 
superconductivity.  
We thus propose that the spin-triplet superconductivity 
is realized in the two-chain Hubbard model by the 
ring-exchange mechanism.  
We argue that the proposed mechanism may have 
possible relevance to triplet superconductivity in 
(TMTSF)$_2$X.  

The minimum model to include the effects of interchain 
coupling may be the two-chain Hubbard model (see 
Fig.~1(b)) defined by the Hamiltonian 
\begin{eqnarray}
{\cal H}=\sum_{<ij>\sigma}t_{ij}
(c^\dagger_{i\sigma}c_{j\sigma}+{\rm H.c.})
+U\sum_i n_{i\uparrow}n_{i\downarrow}
\label{hamiltonian}
\end{eqnarray}
where $c^\dagger_{i\sigma}$ ($c_{j\sigma}$) is the 
creation (annihilation) operator of a hole with spin 
$\sigma$ $(=\uparrow,\downarrow)$ at site $i$ ($j$), 
$n_{i\sigma}=c^\dagger_{i\sigma}c_{i\sigma}$ is the 
number operator, and $\langle ij\rangle$ denotes the 
nearest-neighbor pair.  
We restrict ourselves to the case at quarter filling; 
i.e., $n=\sum_\sigma\langle n_{i\sigma}\rangle=0.5$ 
where $\langle\cdots\rangle$ denotes the ground-state 
expectation value. 
$t_{ij}$ is the hopping integral between sites $i$ 
and $j$: we include $t_1$ along the zigzag chain 
and $t_2$ along the 1D chains.  $U$ is the on-site 
Hubbard repulsion.  
We also examine the intersite repulsive term 
$\sum_{<ij>}V_{ij}n_in_j$ when necessary: we again 
include $V_1$ along the zigzag chain and $V_2$ 
along the 1D chains.  
We hereafter assume $t_2=-1$ as the unit of energy 
$|t_2|=1$; observed small dimerization of hopping 
integrals along the 1D chains is neglected because 
the system remains metallic when the interchain 
hopping integrals are of the zigzag type \cite{nishimoto0}.  
We use values $t_1=\pm 0.25$ and $\pm 0.5$ with the 
sign alternation.  The signs of $t_1$ and $t_2$ 
can instead be taken all positive because the models 
where the product of the three hopping integrals of 
the triangles is positive are equivalent under canonical 
transformation.  
We use a value $U=10$ (in some cases $20$) for 
representing strong electron correlations in real 
systems \cite{nishimoto1}.  
The DMRG method is used for calculating the 
ground-state energy and correlation functions for 
clusters of length $L$ (containing $2L$ 
sites) with open boundary condition.  
We use clusters of up to $L=128$ with keeping up to 
$m\simeq 4500$ density-matrix eigenstates; the discarded 
weights are typically of the order $10^{-7}-10^{-6}$ 
to obtain the ground-state energy in the accuracy 
of $\sim$$0.001|t_2|$.  
Lanczos exact-diagonalization method on small 
clusters is also used.  

\begin{figure}[t]
\begin{center}
\vspace{5pt}
\includegraphics[width=8.0cm,clip]{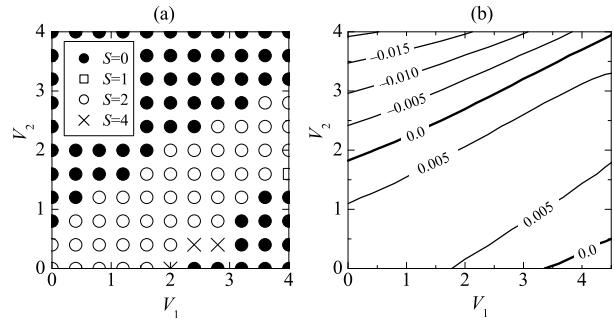}
\vspace{-5pt}
\caption{
(a) Total-spin quantum numbers $S$ of the ground state 
of the $L=8$ cluster with $U=30$ and $|t_1|=1$ plotted 
on the $(V_1,V_2)$ plane.  
(b) Contour plot of the spin correlation 
$\langle{\bf S}_i\cdot{\bf S}_j\rangle$ between the 
interchain nearest-neighbor sites for the $L=8$ cluster 
with $U=10$ and $|t_1|=0.5$ where ${\bf S}_i$ is 
the spin operator of a hole at site $i$.  The 
ground state has $S=0$.}
\end{center}
\label{fig2}
\end{figure}
Basic features of the ground state of our model are 
the following.  
First, the presence of ferromagnetic spin correlation 
is evident in Fig.~2, where we show the phase diagram 
on the $(V_1,V_2)$ plane of an $L$=$8$ cluster with 
periodic boundary condition obtained by an 
exact-diagonalization method.  
We find that when $U$ is large enough the system is 
spin polarized with the total spin $S>0$ (see Fig.~2(a)).  
When $U$ is not large the total spin of the ground state 
becomes $S=0$ but the interchain nearest-neighbor spin 
correlation is still ferromagnetic (see Fig.~2(b)); 
these results are confirmed to persist in larger size 
systems.  
The mechanism is apparent already in the two-hole 
ground state of a three-site Hubbard ring (which is 
spin triplet in a wide parameter region when 
$t_1t_2t_3>0$) and persists in higher dimensions 
as well \cite{penc}.  
We also calculate the charge gap defined as 
$\Delta_c=\lim_{L\rightarrow\infty}\Delta_c(L)$ 
with $\Delta_{\rm c}(L)=E_L(N+2)+E_L(N-2)-2E_L(N)$ 
where $E_L(N)$ is the ground-state energy of a 
chain of length $L$ with $N$ electrons with equal 
number of $\uparrow$ and $\downarrow$ spins.  We 
use the DMRG method for clusters of up to $L=64$ 
and make a finite-size scaling analysis.  
The obtained results (not shown) are similar to 
the results obtained in Refs.~\cite{seo,nishimoto2,ejima}; 
i.e., the charge gap opens due to charge ordering 
when either $V_1$ or $V_2$ is large, but metallic 
state appears in a wide parameter region around 
$V_1=2V_2$ (see Fig.~4 below for the case $V_1=V_2=0$).  
Note that the region where the ferromagnetic 
correlation is strong agrees well with the metallic 
region where the ring exchange of holes works well.  

To see the behavior of the spin degrees of freedom 
further, we calculate the equal-time spin correlation 
function 
$S(i,j)=\langle S_i^zS_j^z\rangle-\langle S_i^z\rangle
\langle S_j^z\rangle$ where $S_i^z$ is the $z$-component 
of the spin operator of a hole at site $i$.  
We also calculate the spin gap defined as 
$\Delta_s=\lim_{L\rightarrow\infty}\Delta_s(L)$
with $\Delta_s(L)=E_L(N_\uparrow+1,N_\downarrow-1)
-E_L(N_\uparrow,N_\downarrow)$ for 
$N_\uparrow=N_\downarrow=N/2$.  The results are shown 
in Fig.~3, where and hereafter we assume $V_1=V_2=0$ 
unless otherwise stated.  We first find that the values 
of $\Delta_s(L)$ are rather small and extrapolated to 
$\Delta_s(L)\rightarrow 0$ within our numerical 
accuracy.  Thus, the spin gap $\Delta_s$ vanishes 
(or becomes quite small if it exists) in the 
thermodynamic limit.  
We also find that the spin correlation decays with 
nearly exponential length dependence, which is 
associated with the oscillations of a period of 4 
times lattice constant (consistent with the period 
of observed $2k_{\rm F}$-SDW state in 
(TMTSF)$_2$X \cite{review}).  
We should note that the apparent contradiction between 
the vanishing spin gap and exponential decay of the 
spin correlation (as well as its rapid decrease at 
$|i-j|\lesssim 30$) may be reconciled by the spin-triplet 
pairing present in our model (see below) \cite{momoi}.  
We point out that the Tomonaga-Luttinger-liquid 
description \cite{solyom} of our results obtained 
in the situation where there are four Fermi points in 
the strong coupling regime is yet an open issue \cite{daul}.  
\begin{figure}[t]
\begin{center}
\vspace{5pt}
\includegraphics[width=6.5cm,clip]{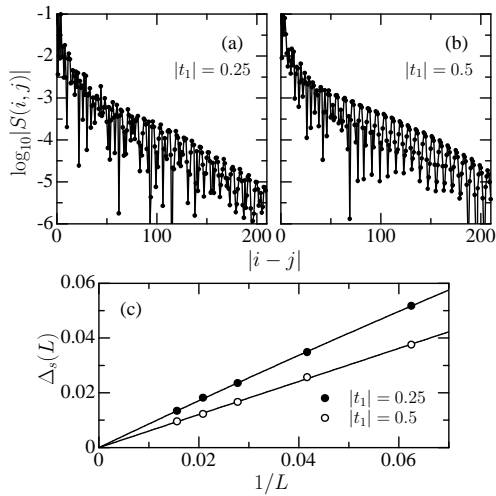}
\vspace{-5pt}
\caption{Spin correlation function 
(a) at $U=10$ and $|t_1|=0.25$ and 
(b) at $U=10$ and $|t_1|=0.5$ for the $L=128$ cluster.  
Horizontal axis $|i-j|$ counts the sites 
along the zigzag chain.  
(c) Spin gap $\Delta_s(L)$ at $U=10$ plotted as 
a function of $1/L$.}
\end{center}
\label{fig3}
\end{figure}

We now calculate the binding energy of holes defined as 
$\Delta_b=\lim_{L\rightarrow\infty}\Delta_b^\pm(L)$ 
with $\Delta_b^\pm(L)=E_L(N\pm 2)+E_L(N)-2E_L(N\pm 1)$.  
Because $E_L(N)$ depends only on $n=N/(2L)$ for a given 
system of length $L$, we have the scaling property 
for large $L$ as 
\begin{eqnarray}
&&\Delta_c(L)={1\over{L}}
{{\partial^2(E_L/L)}\over{\partial n^2}}
+{\cal O}\big({1\over L^3}\big)\\
&&\Delta_b^\pm(L)={1\over{4L}}
{{\partial^2(E_L/L)}\over{\partial n^2}}
+{\cal O}\big({1\over L^3}\big)+\Delta_b
\label{scaling}
\end{eqnarray}
where the proportionality coefficient of $1/L$ is 
related to the inverse of compressibility $\kappa$ 
as 
\begin{equation}
\kappa^{-1}=n^2\lim_{L\rightarrow\infty}
{{\partial^2(E_L/L)}\over{\partial n^2}} .
\end{equation} 
The binding energy $\Delta_b$ in the thermodynamic limit 
is obtained in Eq.~(3) as the ${\cal O}(1)$ correction in 
$\Delta_b^\pm(L)$ \cite{fano}.  
The ${\cal O}(1)$ correction in $\Delta_c(L)$, i.e., 
$\Delta_c$, vanishes when the system is metallic.  

The calculated results are shown in Fig.~4.  
We first note that the values of $\kappa$ 
evaluated from the calculations of $\Delta_c(L)$ and 
$\Delta_b^\pm(L)$ agree well with each other as the 
gradients of the two curves (if the factor 4 is 
taken into account) for small $1/L$ regions agree well.  
Also noted is that the gradient is positive, 
or $\kappa>0$, indicating that the system is 
thermodynamically stable against phase separation.  
We then find that the extrapolated value $\Delta_b$ 
is negative; i.e., the attractive interaction acts 
between holes in the thermodynamic limit.  
The energy gain responsible for the negative value of 
$\Delta_b$ may come from the motion of two holes 
running around the triangle, avoiding the on-site 
repulsion $U$ and exchanging spins for triplet 
coupling, i.e., from the ring exchange of holes.  
The pairing mechanism is thus kinetic in origin.  
\begin{figure}[t]
\begin{center}
\vspace{5pt}
\includegraphics[width=8.0cm,clip]{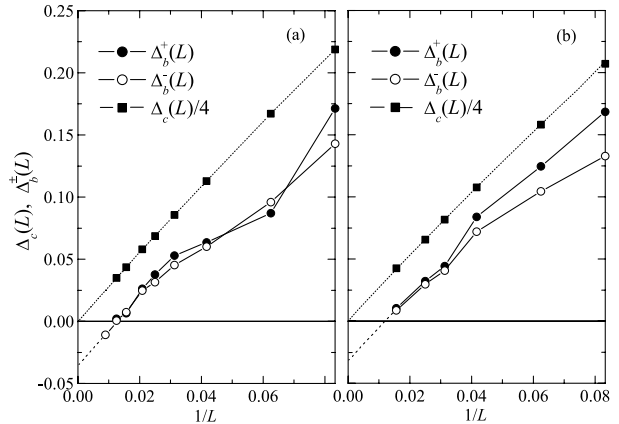}
\vspace{-5pt}
\caption{Binding energy $\Delta_b^\pm(L)$ 
and charge gap $\Delta_c(L)/4$ 
plotted as a function of $1/L$.  
(a) $U=20$ and $|t_1|=0.5$, and 
(b) $U=10$ and $|t_1|=0.5$.  
Solid curve connecting $\Delta_c(L)$ is the 
fitting by a polynomial of $1/L$.  
Solid and dotted lines connecting $\Delta_b^\pm(L)$ 
are a guide to the eyes.}
\end{center}
\label{fig4}
\end{figure}

We also calculate the pair correlation function defined 
as $D(l)=\langle\Delta_{i+l}^\dagger\Delta_i\rangle$
with $\Delta_i=c_{i\uparrow}c_{i+r\downarrow}
-c_{i\downarrow}c_{i+r\uparrow}$ for singlet pairs and 
$\Delta_i=c_{i\uparrow}c_{i+r\downarrow}
+c_{i\downarrow}c_{i+r\uparrow}$ for triplet pairs 
where $i+r$ denotes the neighboring sites of $i$.  
The results at $L=128$ are shown in 
Fig.~5.  We find that $D(l)$ shows the power-law 
length dependence for the interchain triplet pairing 
but decays exponentially for the singlet pairing as 
well as for the triplet pairing on the single 1D chain.  
These are the case also at $L=64$, indicating that the 
size of the clusters used is sufficiently large.  
Quantitatively, the pair correlation function 
(see Fig.~5(b)) at long distances $r_{ij}$ decays 
as $\sim{1/r_{ij}^{1.8}}$, the decay of which is 
slower by comparison than the decay of the charge 
correlation function $\sim{1/r_{ij}^{2.0}}$; the 
estimation is made by fitting the data at $1\ll l\ll L$ 
in order to avoid the effects of the edges of the clusters.  
Then, combined with the attractive interactions of 
two holes shown above, our results indicate that 
the system should be in the state of spin-triplet 
superconductivity where the pairing of holes occurs 
between the interchain nearest-neighbor sites.  
\begin{figure}[t]
\begin{center}
\vspace{5pt}
\includegraphics[width=7.5cm,clip]{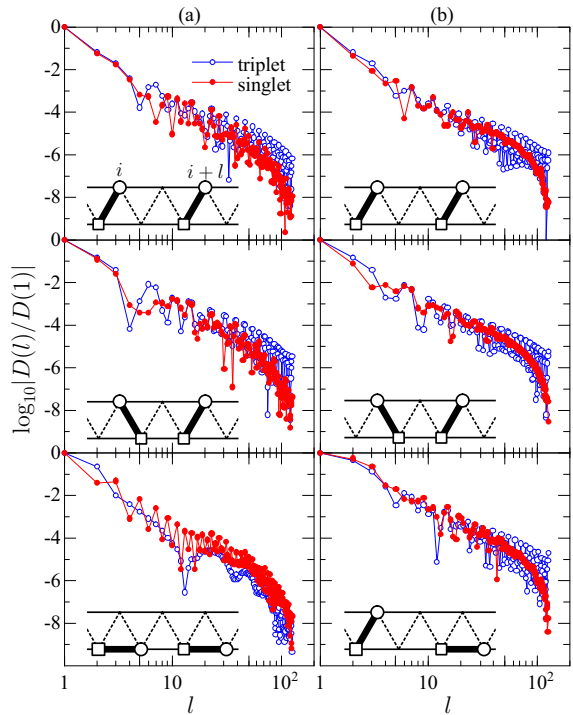}
\vspace{-5pt}
\caption{(Color online) Pair correlation functions $D(l)$ 
calculated at $L=128$ and $U=10$ with 
(a) $|t_1|=0.25$ and (b) $|t_1|=0.5$.  
Horizontal axis $l$ counts the sites along the 
1D chain.  In the inset, the two circles (squares) 
denotes the sites $i$ and $i+l$ ($i+r$ and $i+l+r$), 
where $l$=1 corresponds to the position for two 
squares to sit on the same site.}
\end{center}
\label{fig5}
\end{figure}

Finally, let us consider possible relevance of 
our results to Bechgaard salts.  The model is a 
coupled sequence of the 1D chains.  We have cut 
out two neighboring chains as a minimum model 
to seek for consequences of the interchain 
coupling.  
Since the ring-exchange mechanism works also for 
models of more than two chains \cite{penc} where 
the hopping integrals of all the triangles satisfy 
the ferromagnetic sign rule, we point out that the 
triplet pairing state obtained in the two-chain 
model may persist also in quasi-1D (or 2D) 
systems; future studies will be interesting.  
Also pointed out is that there is intriguing 
competition between the SDW and spin-triplet 
superconducting states in the experimental 
pressure-temperature phase diagram 
\cite{jerome2}.  
Because small intersite Coulombic repulsions exist 
in the materials \cite{nishimoto1}, we examine their 
effects; we find that the inclusion of a realistic 
value $V_2\simeq 1$ \cite{nishimoto1} does not change 
our results.  However, if we include $V_1$ between 
the 1D chains as well, the triplet pair correlation 
becomes less long-ranged and thus the SDW correlation 
can be comparable with it.  A true long-range order 
may be selected among these competing correlations 
when the two (or three) dimensionality is taken into 
account, although to predict which order is realized 
is beyond the scope of the present work.  
We hope that future quantitative analyses will 
help clarifying the issue.  

In summary, we have calculated the binding energy of 
holes and pair correlation functions in the two-chain 
zigzag-bond Hubbard model and have shown that the model 
can be superconducting in spin-triplet channel when 
we make an appropriate choice of the signs of hopping 
integrals for the ferromagnetic ring-exchange 
mechanism to work.  
We have argued that the mechanism proposed may have 
possible relevance to the triplet superconductivity 
in (TMTSF)$_2$X.  

\acknowledgments
We thank G. I. Japaridze, R. M. Noack, K. Sano, Y. Suzumura, 
and M. Tsuchiizu for useful discussions.  
This work was supported in part by Grants-in-Aid for 
Scientific Research from the Ministry of Education, 
Science, Sports, and Culture of Japan. 
A part of computations was carried out at the Research 
Center for Computational Science, Okazaki Research 
Facilities, and the Institute for Solid State Physics, 
University of Tokyo.

\end{document}